\documentclass{PoS}

\title{New Results on Hard Probes in Heavy-Ion Collisions with ALICE}

\ShortTitle{Hard Probes with ALICE}

\author{\speaker{Christian Klein-B\"osing}\thanks{For the ALICE Collaboration and dedicated to Oliver Busch (1976--2018).}\\
        Institut f\"ur Kernphysik, Westf\"alische Wilhelms-Universit\"at M\"unster, Wilhelm-Klemm-Str. 9, 48149 M\"unster, Germany\\
        ExtreMe Matter Institute EMMI, 
        GSI Helmholtzzentrum f\"ur Schwerionenforschung, 
        Planckstr. 1, 
        64291 Darmstadt, 
        Germany\\ 
        E-mail: \email{Christian.Klein-Boesing@cern.ch}}

\abstract{Hard probes -- final state particles related to an interaction with large momentum transfer or mass scale -- play a distinguished role
	in the discovery and the study of the Quark-Gluon Plasma (QGP), a phase of deconfined quarks and gluons reached at high temperatures in heavy ion collisions.
	In heavy ion collisions, parton scatterings with a large momentum transfer ($1/Q \ll 1$\,fm/$c$) occur prior to QGP formation and thus provide a source of coloured probe particles for the QGP created in the later stage of the reaction. 
	The hard scattered partons and the subsequent parton shower interact strongly with the QGP and its constituents via elastic and radiative processes before hadronization into jets of observable particles. 
	Thus, the comparison to jet and high-$\pt$ observables in pp (\emph{vacuum}) potentially probes their modification due to medium effects. 

	One of the key observables in the discovery and investigation of these jet modifications has been the nuclear modification factor $\raa$, for which new results on charged particle production in different colliding systems are presented and the question of apparant suppression in peripheral \pbpb\ collisions is addressed.
	For more differential studies of the jet sub-structure and hence the parton shower evolution in the medium, recent results on jet grooming in heavy ion colisions are presented. 

}

\FullConference{Sixth Annual Conference on Large Hadron Collider Physics (LHCP2018)\\
                4-9 June 2018\\
                Bologna, Italy}

\newcommand{\pt}{\ensuremath{p_\mathrm{T}}}
\newcommand{\raa}{\ensuremath{R_{AA}}}

\newcommand{\ncoll}{\ensuremath{N_\mathrm{coll}}}

\newcommand{\taa}{\ensuremath{T_{AA}}}
\newcommand{\rmd}{\ensuremath{\mathrm{d}}}
\newcommand{\snn}{\ensuremath{\sqrt{s_\mathrm{NN}}}}

\newcommand{\pbpb}{Pb-Pb}
\newcommand{\xexe}{Xe-Xe}
\newcommand{\ppb}{p-Pb}

\begin{document}


\section{Introduction}

%
%
%

The nuclear modification factor $\raa$ compares the particle production in nucleus-nucleus reaction ($AA$) to pp collisions scaled with the number of binary collision ($\ncoll$) or the increased parton flux ($\taa$) in nuclear collisions
\begin{equation}
\raa = \frac{\rmd N_{X}^{AA}/\rmd\pt}{\ncoll \cdot \rmd N_{X}^{pp}/\rmd\pt} = \frac{\rmd N_{X}^{AA}/\rmd\pt}{\taa \cdot \rmd \sigma_{X}^{pp}/\rmd\pt},
\end{equation}
where $X$ can be any reconstructed final state object: charged particles, specific hadrons, leptons, gauge bosons or jets.
In the absence of any strong  medium effects in the initial and final state, $\raa$ is expected to be unity for single particles in the $\pt$  region where hard processes are the dominant source of particle production. 
%
%
%

For reconstructed jets, the interpretation and comparison of $\raa$ measurements is more sophisticated. 
In principle, a jet algorithm aims to recover the full kinematic information on the initial parton.
 If this also holds in $AA$ collisions, one would expect a $\raa$ of unity, i.e. the full energy is recovered and all medium modification is visible only in a change of the jet structure.
In practice, jet reconstruction in heavy ion collisions is hindered by the large soft background, unrelated to the initial hard scatterings.
This background can be subtracted with an event-wise average of the background density and is in general reduced for jets with smaller radius and larger constituent-$\pt$ thresholds.
%
%
Different experimental choices on $\pt$ thresholds and jet radii, together with given detector-specific methods for background corrections and input for jet reconstruction, make a direct comparison between different experiments difficult. 
However, in jet $\raa$ measurements a general trend is observed by all LHC experiments: the nuclear modification factor in central collisions appears to approach an asymptotic value of $\approx 0.5$ at high $\pt$ similar to the measurements of single charged hadrons.
%


The data presented here have been collected with the ALICE experiment during various runs of the Large Hadron Collider at CERN. 
In particular, the precise measurement of charged particle tracks at central (pseudo-)rapidity ($\left|\eta\right|< 0.9$) in high multiplicity environment and down to low $\pt \approx 0.15$\,GeV allows for a detailed quantification of the jet structure and the characterisation of the underlying background~\cite{Abelev:2014ffa}.
%
%
%
%
%
%


\section{System Size Dependence and Centrality Biases}

\begin{figure}[th]
	\unitlength\textwidth
	\begin{picture}(1,0.39)
	\put(0.1,0.00){\includegraphics[width=.3\textwidth]{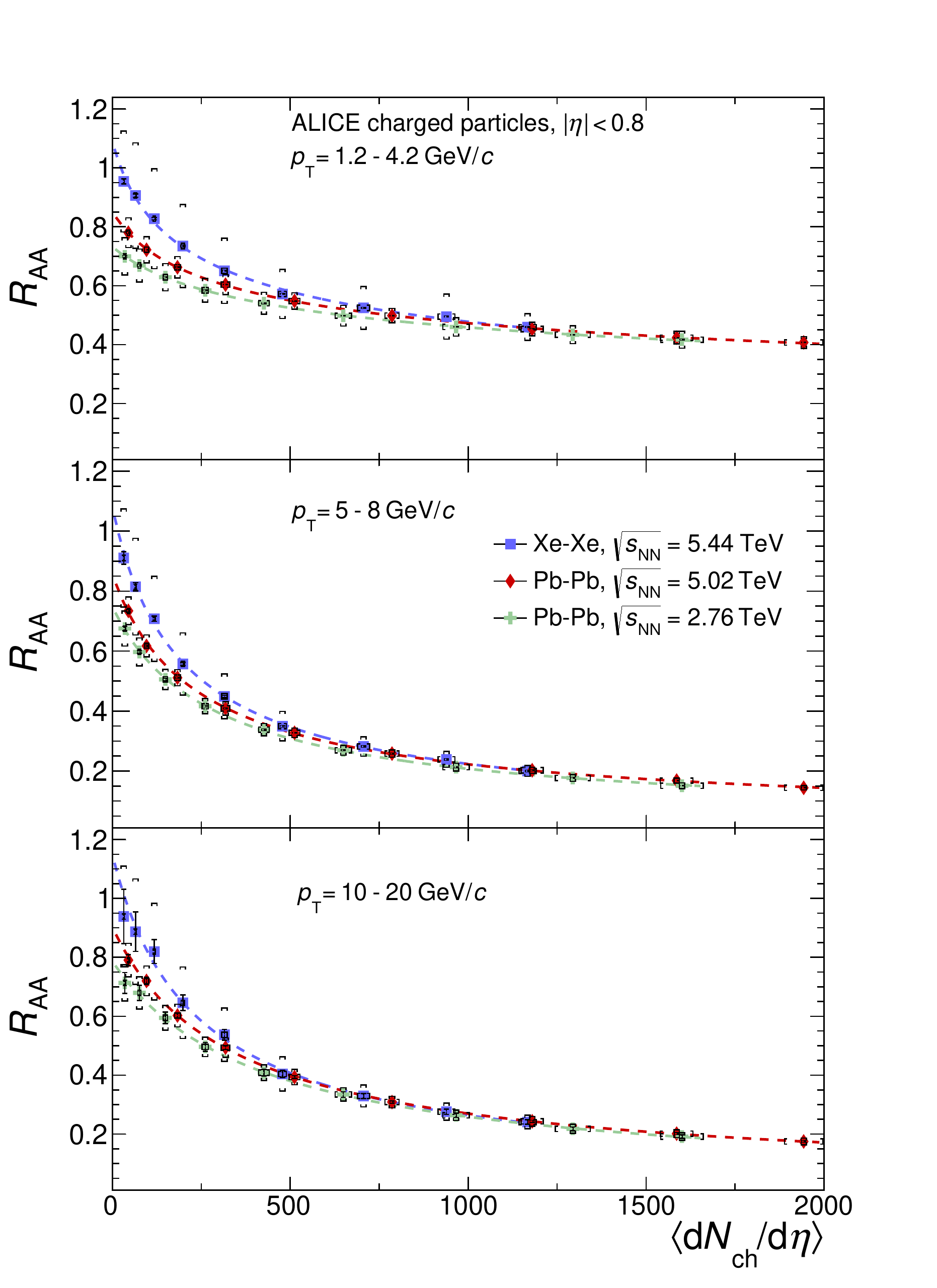}}
	\put(0.5,0.00){\includegraphics[width=.4\textwidth]{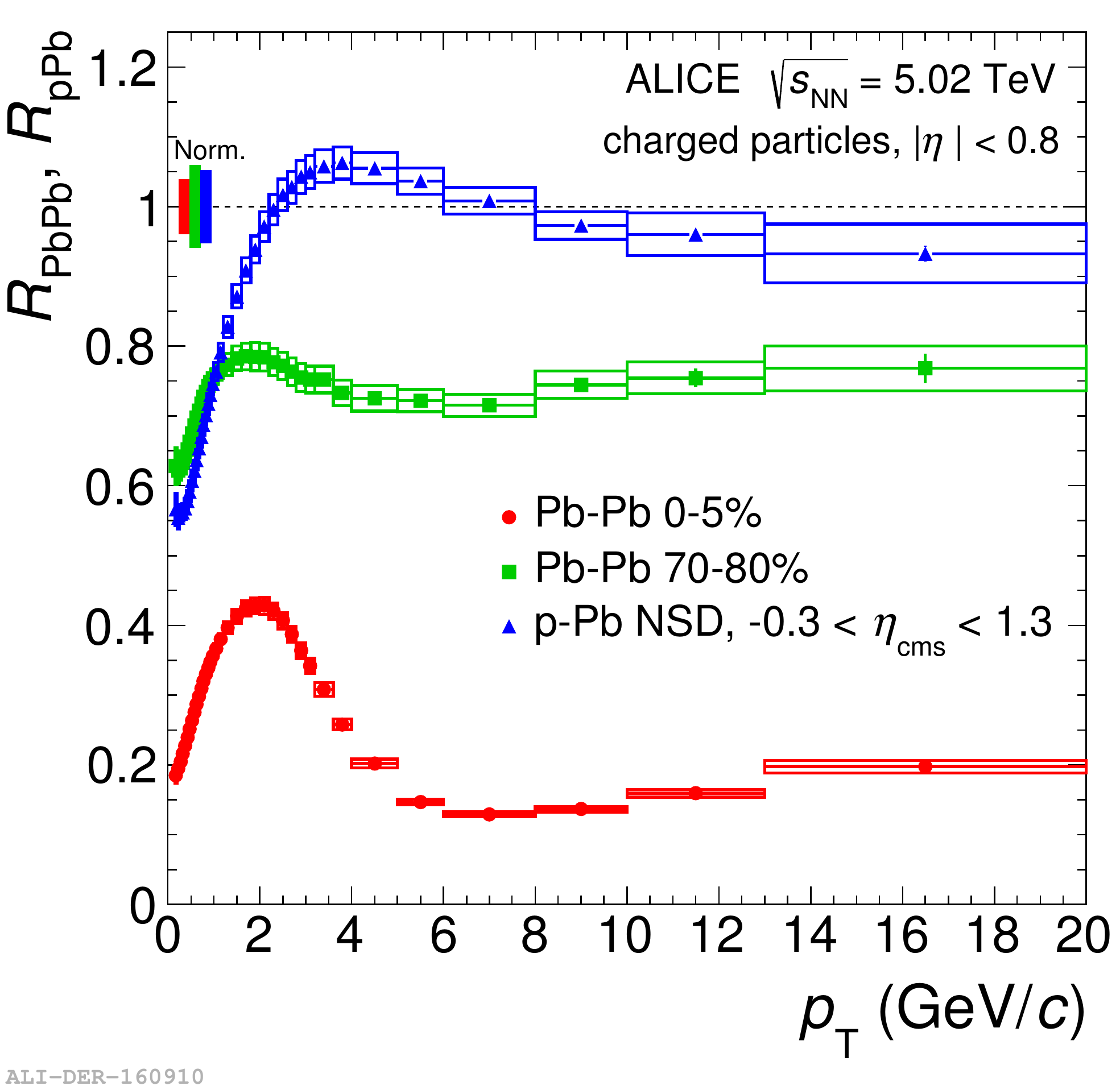}}
	\end{picture}
	\caption{ Left: Nuclear modification factor for charged particles in three $\pt$ ranges for  different colliding systems and energies \cite{Acharya:2018eaq} Right: Nuclear modification in \ppb\ as well as in peripheral and central \pbpb\ \cite{Acharya:2018qsh,ALICE:2012mj}}
	\label{fig8-10}
\end{figure}

In the measurement of single charged particles at high $\pt$ the ALICE collaboration recently published the results for \pbpb\ collisions at the highest collision energy so far, $\snn = 5.02$ TeV, with significantly improved systematic uncertainties and a $\pt$ reach up to 50 GeV \cite{Acharya:2018qsh}. 
The nuclear modification factor $\raa$ shows only little variation from the measurements at $\snn = 2.76$ TeV, which can be understood when considering that the expected larger parton energy loss is compensated by a harder parton spectrum by which the energy loss is filtered.
The harder spectrum at LHC leads in general to an increased importance of the subleading fragments already in the single particle spectrum compared to RHIC. 
This directly affects the description of $\raa$, where parton energy loss models beyond leading particles perform better at the LHC \cite{Abelev:2012hxa}. 
In addition, it is already visible in the comparison of high-$\pt$ identified particles in pp collisions at large $\sqrt{s}$ that leading order and next-to-leading order (LO and NLO) Monte Carlo Models with parton showering provide a better description of the data than NLO perturbative QCD calculation with one dimensional fragmentation functions \cite{Acharya:2017tlv} extracted at a lower collisions energy.

\begin{figure}
	\unitlength\textwidth
	\begin{picture}(1,0.33)
	\put(0.04,0.00){\includegraphics[width=.45\textwidth]{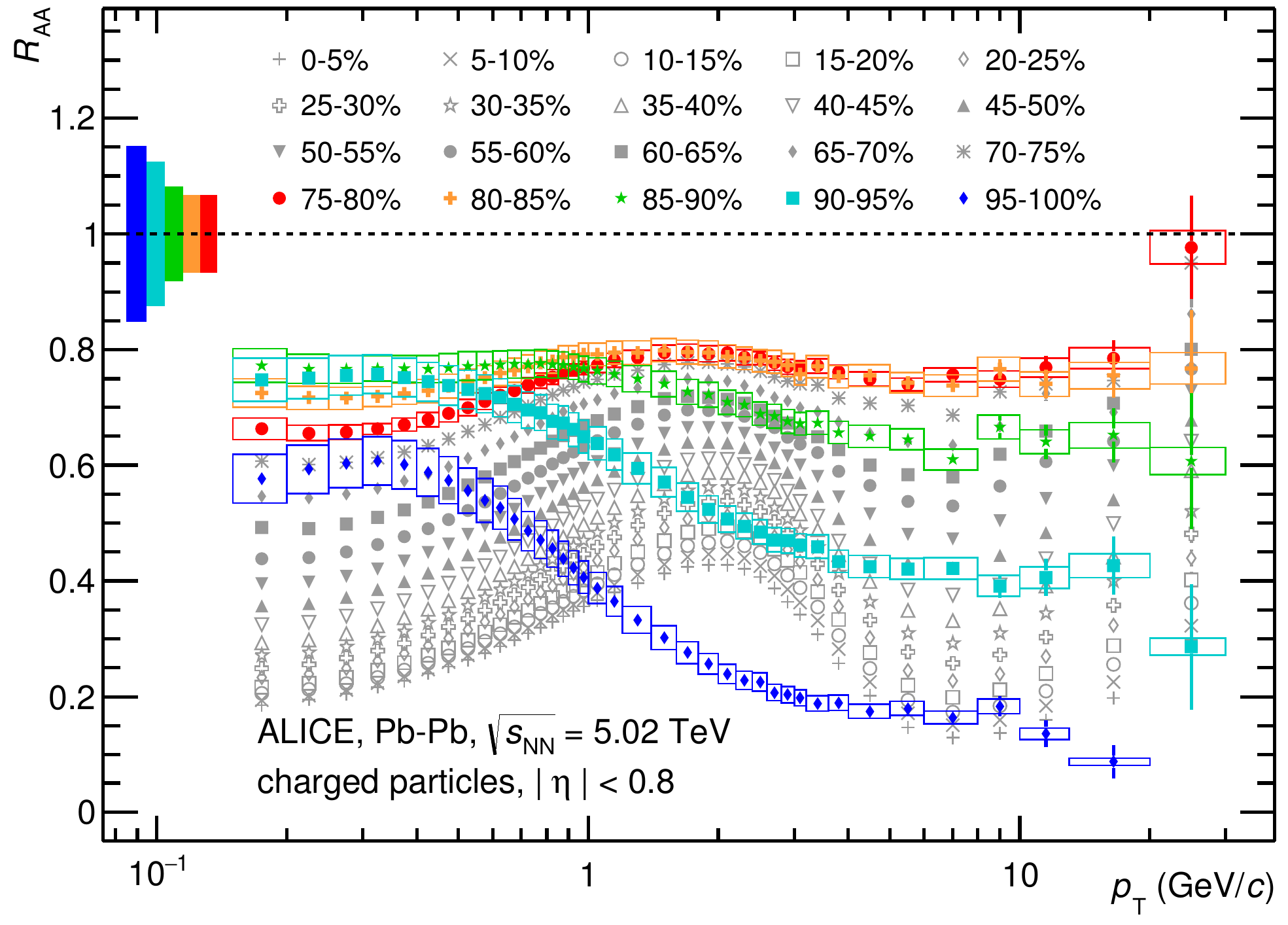}}
	\put(0.58,0.){\includegraphics[width=.33\textwidth]{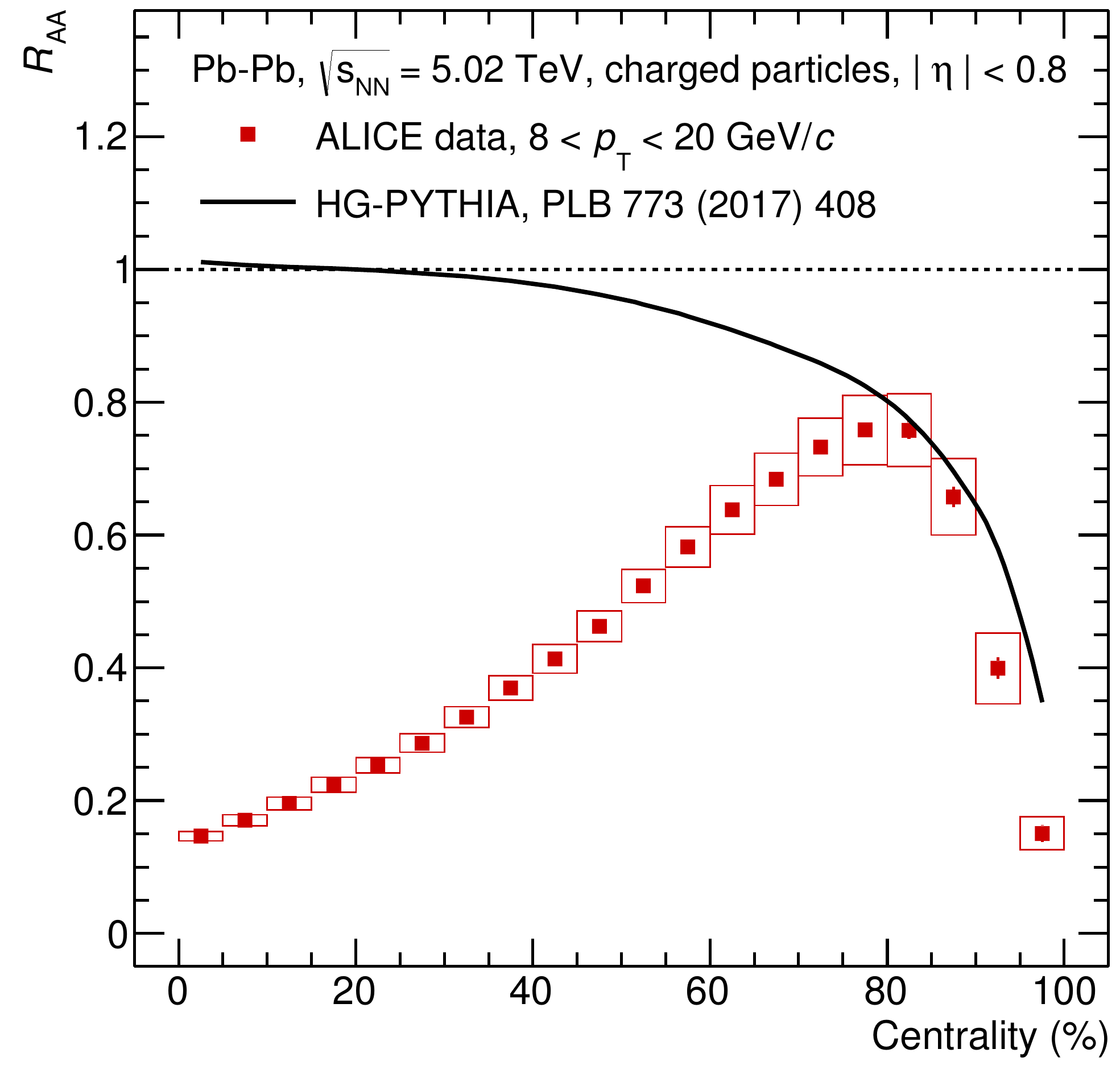}}
	\end{picture}
	\caption{Left: Nuclear modification factor for charged particles over the full centrality range \cite{Acharya:2018njl} Right: Comparison of the apparent suppression to the expectation of the \textsc{Hg-Pythia} model \cite{Morsch:2017brb}}
	\label{fig11}
\end{figure}

The comparison of the charged particle production in \pbpb\ at $\snn = 5.02$ TeV \cite{Acharya:2018qsh} to recent ALICE results obtained with lighter Xe-ions at similar energies ($\snn = 5.44$ TeV, \cite{Acharya:2018eaq}) is important to disentangle effects of the collision geometry.
In particular this can be done by studying the nuclear modification factor in three distinct $\pt$ ranges as shown in Fig.\,\ref{fig8-10} (left). 
The lowest $\pt$ range is dominated by effects induced by the collective expansion of the medium (flow).
In the intermediate range, pronounced differences between baryon and mesons have been observed, which could be explained by quark coalescence, while in the highest $\pt$ range above 10\,GeV the effects of parton energy loss dominate.
Plotted as function of the charged particle multiplicity as measure for the energy density, it is remarkable that all data points agree beyond $\rmd N/\rmd\eta \approx 500$. 
Irrespective of the ion size and colliding energy, the driving mechanism in all three $\pt$ ranges appears to be energy density, even though the ranges are dominated by different physics processes.
At low $\rmd N/\rmd\eta$ the nuclear modification factors for Xe and Pb data deviate.
At the same $\rmd N/\rmd\eta$, \xexe\ collisions are more asymmetric, suggesting that the collision geometry now becomes an additional factor. 
Interestingly, these deviations are similar in all three $\pt$ ranges, which highlights the importance of  geometry as a connection between the various processes involved in particle production in heavy ion collisions.

The search for the system size and/or energy density dependence of parton energy loss can also be viewed from a different perspective, when comparing peripheral \pbpb\ with \ppb\ at similar $\rmd N/\rmd\eta$. 
Indeed, as seen in Fig.\,\ref{fig11} (left), even for peripheral \pbpb\ reactions the nuclear modification factor at high $\pt$ does not increase beyond 0.8 \cite{Acharya:2018njl}.
This apparent contradiction to the observation of $\raa \approx 1$ in \ppb\ in Fig.\,\ref{fig8-10} (right), can be explained when considering that indeed not the number of binary nucleon-nucleon collisions is the relevant scaling assumption, but the number of \emph{hard} collisions. 
This number depends on the probability for multiple parton interactions (MPI), which is not uniform across the nucleus as assumed in geometric Glauber models. 
As discussed in \cite{Morsch:2017brb}, the average distance between two colliding nucleons is larger on average in peripheral collisions compared to central, which leads to fewer MPI per $\ncoll$. 
The geometry bias is further enhanced by the experimental centrality selection on multiplicity, which in turn is increasing with the number of MPIs. 
These biases have been evaluated in a hybrid model, coupling the impact parameter dependent number of MPIs from \textsc{Hijing} with \textsc{Pythia}. 
As shown in  Fig.\,\ref{fig11} (right), the apparent suppression can be very well reproduced in this \textsc{Hg-Pythia}-Model, without the need to invoke any other initial or final state effects.

\begin{figure}
	\unitlength\textwidth
	\begin{picture}(1,0.39)
	\put(0.03,0.01){\includegraphics[width=.55\textwidth]{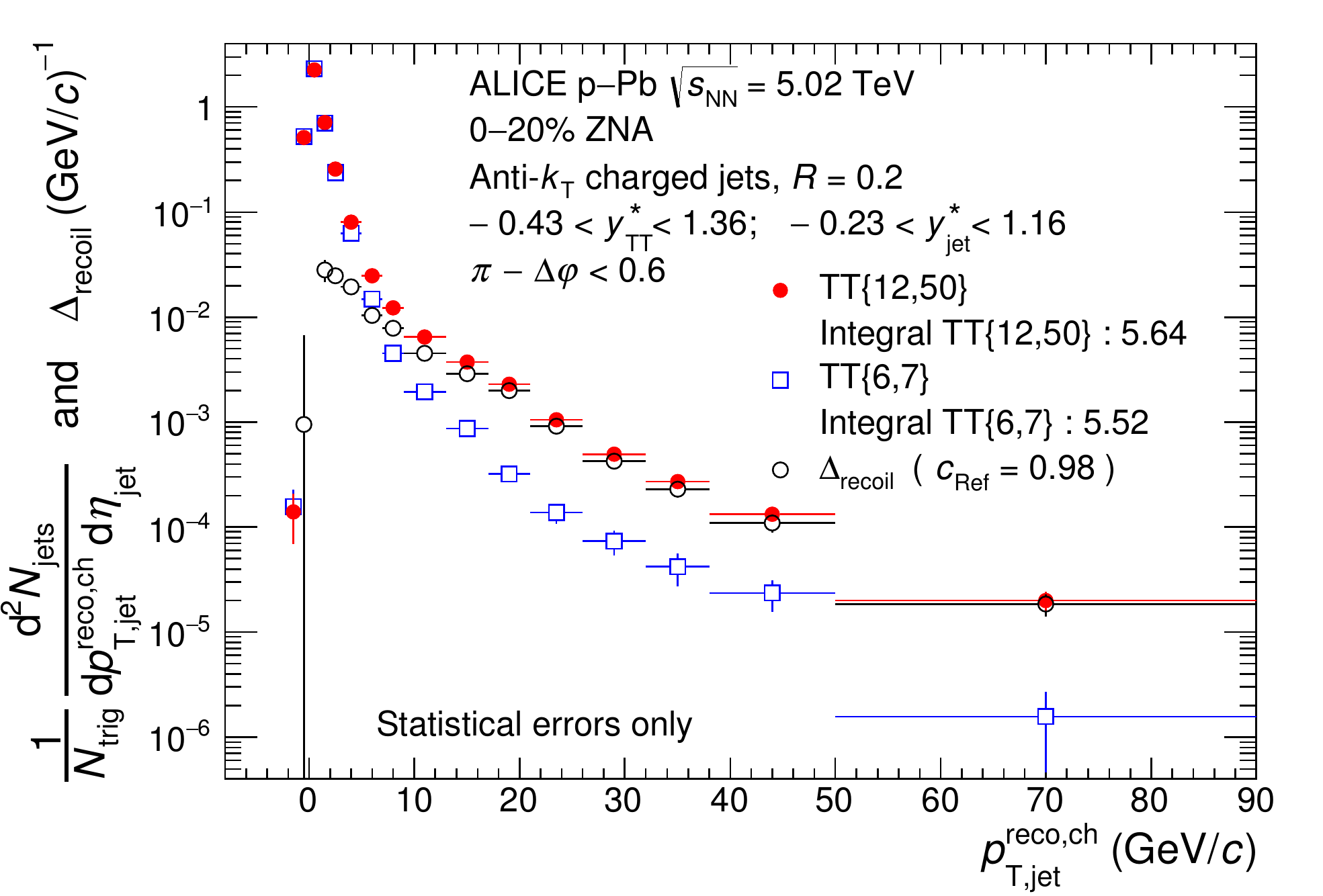}}
	\put(0.58,0.){\includegraphics[width=.4\textwidth]{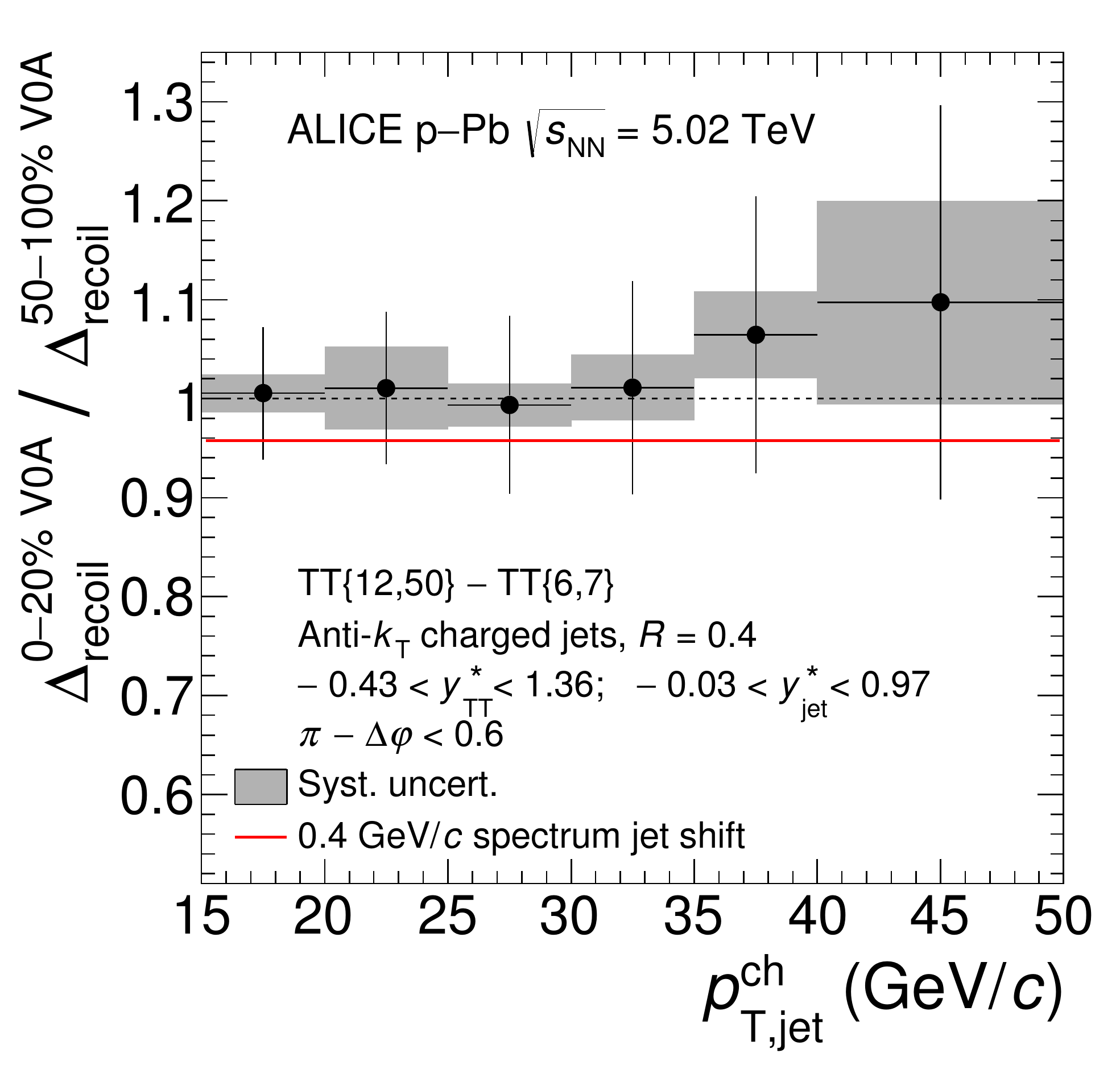}}
	\end{picture}
	\caption{Left: Jet spectra opposite of a trigger hadron with given $\pt$ and difference spectrum between them Right: Comparison of difference spectra in \ppb\ collisions with different event activity \cite{Acharya:2017okq}}
	\label{fig16a}
\end{figure}

An unbiased measure for parton energy loss, or redistribution of jet energy, has been introduced by the ALICE collaboration with the measurement of jet distributions with respect to a recoiling hadron \cite{Adam:2015doa}.
In these measurements a deliberate bias is put on the momentum transfer $Q^2$ in the reaction by requiring a certain hadron trigger $\pt$. 
The jet distribution is reconstructed opposite in $\varphi$ to this trigger and is shown as an example in   
Fig.\,\ref{fig16a} (left): a  clear hardening with increasing trigger $\pt$ is observed.
Now the difference of the spectra can be compared for various multiplicities in \ppb. Since this is a \emph{per-trigger} quantity (i.e. per hard collision) it is independent of the biases discussed above. 
It is seen that the energy loss/spectrum shift in \ppb\ is compatible with no effect and smaller than 0.4 GeV at 90\% confidence level \cite{Acharya:2017okq}.  

\section{Differential Jet Observables}

\begin{figure}
	\unitlength\textwidth
	\begin{picture}(1,0.39)
	\put(0.03,0.03){\includegraphics[width=.55\textwidth]{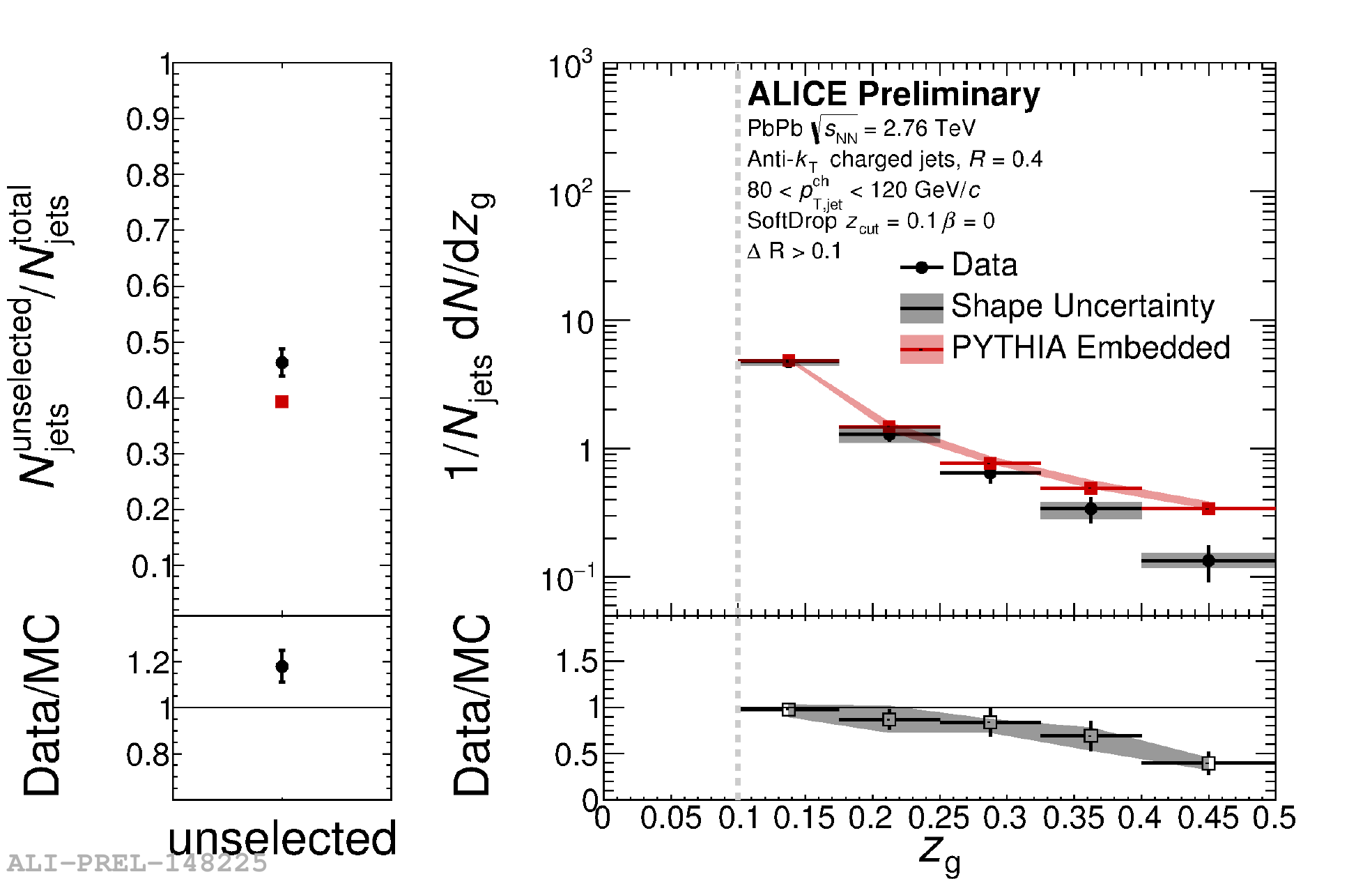}}
	\put(0.6,0.){\includegraphics[width=.4\textwidth]{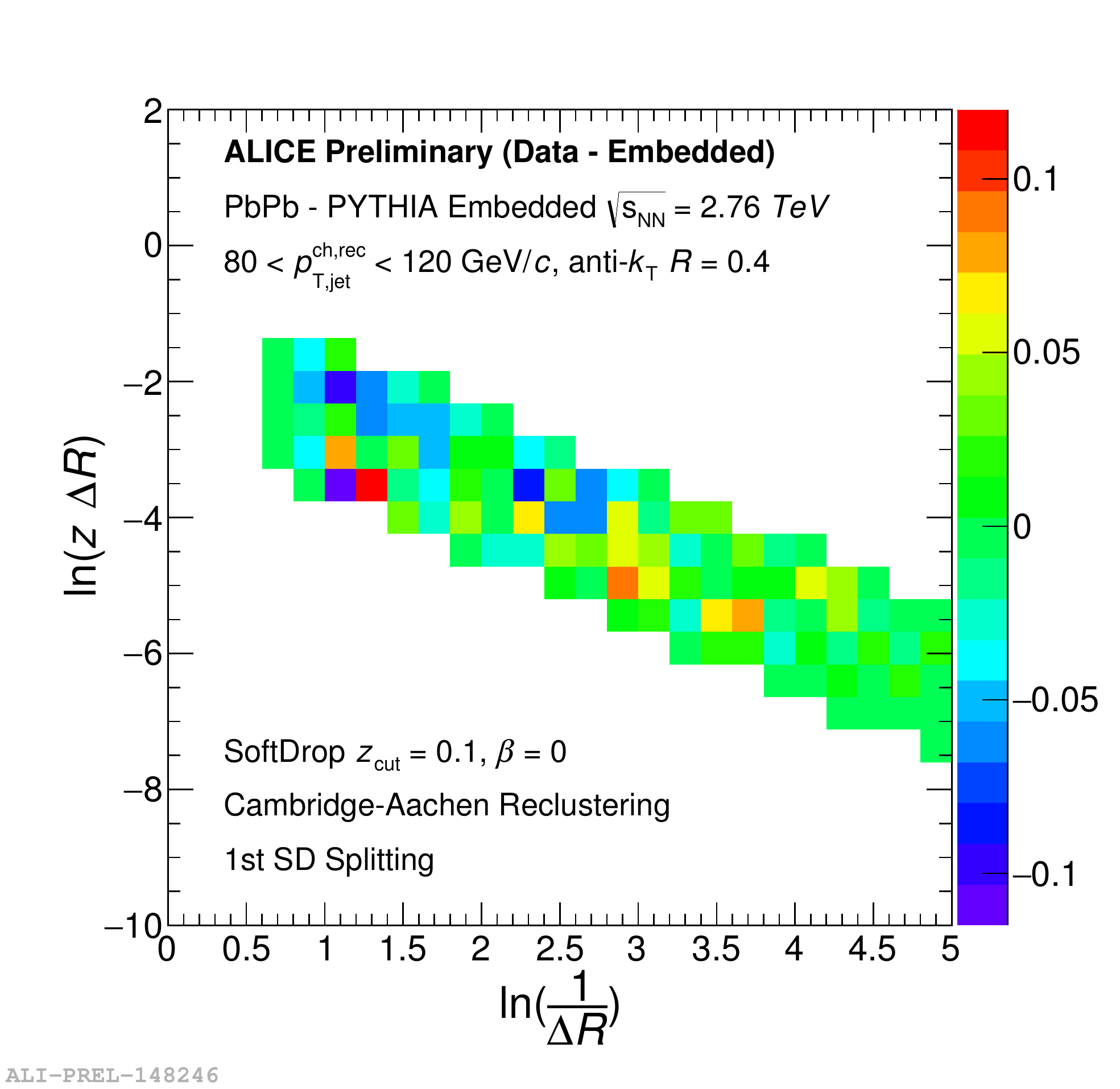}}
	\end{picture}
	\caption{Left: Inclusive measurement of $z_{g}$ distribution in central \pbpb\ collisions at $\sqrt{s}$ = 2.76 TeV with $\Delta R$ cut $> 0.1$ Right: Lund representation of the difference between Pb-Pb data splittings and \textsc{Pythia} embedded level splittings in the 10\% most central collisions at $\sqrt{s}=2.76$ TeV. \cite{Andrews:2018wgw}}
	\label{fig18}
\end{figure}

The differential study of jet structure has recently seen the development of a wealth of observables, 
well beyond cross sections ratios for different radii or transverse momentum distributions of jet constituents. 
One particular development is to facilitate the widely used sequential recombination algorithms to \emph{decluster} a found jet and essentially rewind the QCD splitting.
This can be done e.g. with the Cambridge--Aachen algorithm, which is based on angular proximity only. Each step in the procedure splits a cluster into two sub-clusters (sub-jets), separated by $\Delta R$ and with momentum fractions 
$z$ and $1 -z$, respectively. 
The $z$ and $\Delta R$ values can be used to populate a \emph{Lund diagram}, which in principle maps the phase space of all splittings and allows for the isolation of different regions for medium effects. 
Certain regions of interest can also be amplified or filtered by \emph{grooming} methods, e.g. the \emph{soft drop} \cite{Larkoski:2014wba} algorithm unwinds the jets, following the largest $z$ (dropping the soft sub-jet), until
$z > z_\mathrm{cut} \Delta R^\beta$. 
Preliminary results on the $z$ distribution of the first splitting identified with this procedure are shown in Fig.\,\ref{fig18} (left), using $z_\mathrm{cut} = 0.1, \beta = 0$ in central \pbpb\ collisions. 
A constituent subtraction scheme has been applied to take into account the background (see also \cite{Andrews:2018wgw} and references therein). 
The normalization to the total number of jets and the comparison to \textsc{Pythia} jets embedded into central \pbpb\ events reveals that there is no sign of enhanced large angle $(\Delta R > 0.1)$ splitting at any $z$. 
Further studies in different $\Delta R$ regions \cite{Andrews:2018wgw}, show that there is an indication for overall enhancement of collimated splitting and suppression of large angle splittings. 
This behaviour is also visible in the Lund representation of the difference between \pbpb\ data splittings and \textsc{Pythia} embedded level splittings in  Fig.\,\ref{fig18} (right).
%





\section{Summary and Outlook}

In summary, the $\raa$ observable alone cannot reflect the complexity of jet medium modification but is still highly popular. 
A new detailed picture of the parton shower is provided by the Lund diagram and groomed jets have been used as a tool to select jet substructure.
Both methods show a suppression of large angle splittings and enhancement of collinear splittings compared to embedded \textsc{Pythia}.
These results, together with the wealth of ALICE hard probes and jet measurements building on the inclusion of low $\pt$ constituents, provide an important benchmark for the test and development of  Monte Carlo frameworks coveringr pp, p$A$ and $AA$ reactions, including the modeling of parton shower modification and the underlying event.

This contribution is dedicated to Oliver Busch (1976 -- 2018); a dear friend and long-term collaborator who has driven many jet analysis within ALICE and whose work will have a lasting impact on the jet programme to come. 


\bibliographystyle{JHEP}
\bibliography{master}

\providecommand{\href}[2]{#2}\begingroup\raggedright\begin{thebibliography}{10}

\bibitem{Abelev:2014ffa}
{\scshape ALICE} collaboration, B.~Abelev et~al., \emph{{Performance of the
  ALICE Experiment at the CERN LHC}},
  \href{https://doi.org/10.1142/S0217751X14300440}{\emph{Int. J. Mod. Phys.}
  {\bfseries A29} (2014) 1430044}
  [\href{https://arxiv.org/abs/1402.4476}{{\ttfamily 1402.4476}}].

\bibitem{Acharya:2018eaq}
{\scshape ALICE} collaboration, S.~Acharya et~al., \emph{{Transverse momentum
  spectra and nuclear modification factors of charged particles in Xe-Xe
  collisions at $\sqrt{s_\mathrm{ NN}}$ = 5.44 TeV}},
  \href{https://arxiv.org/abs/1805.04399}{{\ttfamily 1805.04399}}.

\bibitem{Acharya:2018qsh}
{\scshape ALICE} collaboration, S.~Acharya et~al., \emph{{Transverse momentum
  spectra and nuclear modification factors of charged particles in pp, p-Pb and
  Pb-Pb collisions at the LHC}},
  \href{https://arxiv.org/abs/1802.09145}{{\ttfamily 1802.09145}}.

\bibitem{ALICE:2012mj}
{\scshape ALICE} collaboration, B.~Abelev et~al., \emph{{Transverse Momentum
  Distribution and Nuclear Modification Factor of Charged Particles in $p$-Pb
  Collisions at $\sqrt{s_{NN}}=5.02$ TeV}}, {\emph{Phys. Rev. Lett.} {\bfseries
  110} (2013) 082302} [\href{https://arxiv.org/abs/1210.4520}{{\ttfamily
  1210.4520}}].

\bibitem{Abelev:2012hxa}
{\scshape ALICE} collaboration, B.~Abelev et~al., \emph{{Centrality Dependence
  of Charged Particle Production at Large Transverse Momentum in Pb--Pb
  Collisions at $\sqrt{s_\mathrm{{NN}}} = 2.76$ TeV}},
  \href{https://doi.org/10.1016/j.physletb.2013.01.051}{\emph{Phys. Lett.}
  {\bfseries B720} (2013) 52}
  [\href{https://arxiv.org/abs/1208.2711}{{\ttfamily 1208.2711}}].

\bibitem{Acharya:2017tlv}
{\scshape ALICE} collaboration, S.~Acharya et~al., \emph{{$\pi ^{0}$ and $\eta
  $ meson production in proton-proton collisions at $\sqrt{s}=8$ TeV}},
  \href{https://doi.org/10.1140/epjc/s10052-018-5612-8}{\emph{Eur. Phys. J.}
  {\bfseries C78} (2018) 263}
  [\href{https://arxiv.org/abs/1708.08745}{{\ttfamily 1708.08745}}].

\bibitem{Acharya:2018njl}
{\scshape ALICE} collaboration, S.~Acharya et~al., \emph{{Analysis of the
  apparent nuclear modification in peripheral Pb-Pb collisions at 5.02 TeV}},
  \href{https://arxiv.org/abs/1805.05212}{{\ttfamily 1805.05212}}.

\bibitem{Morsch:2017brb}
C.~Loizides and A.~Morsch, \emph{{Absence of jet quenching in peripheral
  nucleus--nucleus collisions}},
  \href{https://doi.org/10.1016/j.physletb.2017.09.002}{\emph{Phys. Lett.}
  {\bfseries B773} (2017) 408}
  [\href{https://arxiv.org/abs/1705.08856}{{\ttfamily 1705.08856}}].

\bibitem{Acharya:2017okq}
{\scshape ALICE} collaboration, S.~Acharya et~al., \emph{{Constraints on jet
  quenching in p-Pb collisions at $\mathbf{\sqrt{s_{NN}}}$ = 5.02 TeV measured
  by the event-activity dependence of semi-inclusive hadron-jet
  distributions}},
  \href{https://doi.org/10.1016/j.physletb.2018.05.059}{\emph{Phys. Lett.}
  {\bfseries B783} (2018) 95}
  [\href{https://arxiv.org/abs/1712.05603}{{\ttfamily 1712.05603}}].

\bibitem{Adam:2015doa}
{\scshape ALICE} collaboration, J.~Adam et~al., \emph{{Measurement of jet
  quenching with semi-inclusive hadron-jet distributions in central Pb-Pb
  collisions at ${\sqrt{{s}_{\mathrm{NN}}}}$ = 2.76 TeV}},
  \href{https://doi.org/10.1007/JHEP09(2015)170}{\emph{JHEP} {\bfseries 09}
  (2015) 170} [\href{https://arxiv.org/abs/1506.03984}{{\ttfamily
  1506.03984}}].

\bibitem{Larkoski:2014wba}
A.~J. Larkoski, S.~Marzani, G.~Soyez and J.~Thaler, \emph{{Soft Drop}},
  \href{https://doi.org/10.1007/JHEP05(2014)146}{\emph{JHEP} {\bfseries 05}
  (2014) 146} [\href{https://arxiv.org/abs/1402.2657}{{\ttfamily 1402.2657}}].

\bibitem{Andrews:2018wgw}
{\scshape ALICE} collaboration, H.~A. Andrews, \emph{{Exploring the Phase Space
  of Jet Splittings at ALICE using Grooming and Recursive Techniques}},
  {\emph{Submitted to: Nucl. Phys.} (2018) }
  [\href{https://arxiv.org/abs/1807.06439}{{\ttfamily 1807.06439}}].

\end{thebibliography}\endgroup

\end{document}